\documentclass[prl,aps,twocolumn]{revtex4}
\usepackage{amsfonts}
\usepackage{amssymb}
\usepackage{graphicx}
\usepackage{mathrsfs}
\usepackage{array}
\usepackage{amsmath}
\usepackage{lscape}
\usepackage{ bbold }
\usepackage{ucs}

\begin{document}

\title{Transitions between spatial attractors in place-cell networks}

\author{R. Monasson, S. Rosay}

\affiliation{Laboratoire de Physique Th\'eorique de l'ENS, CNRS \& UPMC, 24 rue Lhomond,75005 Paris, France}

\date{\today}

\begin{abstract}
The spontaneous transitions between $D$-dimensional spatial maps in an attractor neural network are studied. Two scenarios for the transition from on map to another are found, depending on the level of noise: (1) through a mixed state, partly localized in both maps around positions where the maps are most similar; (2) through a weakly localized state in one of the two maps, followed by a condensation in the arrival map. Our predictions are confirmed by numerical simulations, and qualitatively compared to recent recordings of hippocampal place cells during quick-environment-changing experiments in rats.
\end{abstract}
\maketitle


In 1982 J.J. Hopfield proposed a neural network model for auto-associative memories, in which specific configurations (patterns) of the activity could be stored through an adequate choice of the interactions between the neurons, modeled as binary units $\sigma=0,1$ \cite{Hopfield82}. Starting from an initial configuration partially ressembling one pattern the network configuration dynamically evolves until a fixed point, coinciding almost exactly with the stored pattern, is reached \cite{AGS85}. In this paper we consider an extension of the Hopfield model, in which the attractors stored are continuous and $D$-dimensional (with $D\ge 1$), rather than discrete fixed points ($D=0$), and discuss the existence and the mechanisms of spontaneous transitions between those attractors. Besides its intrinsic interest from a statistical mechanics point of view our study is motivated by the observation of abrupt transitions between the representations of space in the brain \cite{Wills05}, in particular in quick-environment-changing  experiments on rats \cite{Jezek11}. 

Continuous attractors are not unusual in statistical physics. An illustration is given by the Lebowitz-Penrose theory of the liquid-vapor transition \cite{Lebowitz66}. Consider a $D$-dimensional lattice, whose $N$ sites $\vec x_i$ can be occupied by a particle ($\sigma_i=1$), or left empty ($\sigma_i=0$). The energy of a configuration $\{\sigma_i\}$ is given by the Ising-like Hamiltonian
\begin{equation}\label{LP}
E\big[ \{\sigma_i\}, \{\vec x_i\}\big] = - \sum_{i<j} J\big(|\vec x_i-\vec x_j|\big) \, \sigma_i \,\sigma_j \ ,
\end{equation}
where $J$ is a positive and decaying function of its argument, {\em i.e.} of the distance between sites. At fixed number of particles and low enough temperature translation invariance on the lattice is spontaneously broken: particles tend to cluster in the $\vec x$-space, and form a high density region (liquid drop) surrounded by a low-density vapor. The density profile of  this `bump' of particles hardly fluctuates, but its position can freely move on the lattice, and defines a collective coordinate for the microscopic configuration of particles.

From the neuroscience point of view, the existence of a collective coordinate, weakly sensitive to the high stochasticity of the microscopic units, is central to population coding theory \cite{Amari77}. Following the seminal discovery of `place cells' in a brain area called hippocampus \cite{OKeefe71}, continuous attractors have been proposed as the basic principle of the coding for position in space \cite{Tsodyks95}. The model we consider here goes as follows \cite{Monasson13}. In a rat moving freely in a given environment, a place cell $i$ becomes active ($\sigma_i=1$) when the rat is at a specific location in the environment, called place field and  centered in $\vec x_i$, and silent otherwise ($\sigma_i=0$). Place cells $i,j$ with overlapping place fields, {\em i.e.} such that the distance between $\vec x_i$ and $\vec x_j$ is small enough, may be simultaneously active, and have a tendency to strengthen their connection during the exploration of the environment. The potentiation of the couplings between coactive neurons is called Hebbian learning, an important paradigm in auto-associative memories. In addition to the local excitation, global inhibition across the network keeps the fraction of active units constant. As a result  the probability of a place-cell activity configuration $\{\sigma_i\}$ would formally coincide with the Gibbs measure associated to model (\ref{LP}), with a temperature $T$ dependent on the neural noise.  
%

When the rat explores a new environment hippocampal place cells undergo a process called remapping, in which place field locations are randomly reallocated \cite{Kubie91}. A simple model of remapping consists in randomly permuting the indices of the place field locations, {\em i.e.} the center of the place field attached to place cell $i$ becomes $\vec x_{\pi(i)}$, where $\pi$ is a random permutation defining the spatial map of the environment. We assume that the contributions to the interactions of the different maps add up, and obtain the Hopfield-like hamiltonian
\begin{equation}\label{ours}
E\big[ \{\sigma_i\}, \{\vec x_i\},\{\pi^\ell\}\big] =- \sum_{i<j} J_{ij} \, \sigma_i \,\sigma_j  \ ,
\end{equation}
where the couplings are
\begin{equation}\label{jij}
J_{ij} =  \sum_{\ell =1}^L J\left(\left|\vec x_{\pi^\ell(i)}-\vec x_{\pi^\ell(j)}\right|\right)   \ , 
\end{equation}
$L$ is the number of environments, and $\pi^\ell$ the permutation in the $\ell^{th}$ environment ($\ell=1,..., L$). This $J_{ij}$-matrix exhibits a `small-world'-like topology \cite{smallworld}: the couplings due to an environment, say, $\ell=\ell_0$ in (\ref{jij}), connect neurons with place fields close to each other in this environment, while the other environments ($\ell \ne \ell_0$) contribute long-range and random connections. Despite this structural disorder, in model (\ref{ours}) and in similar rate-based models with continuous units \cite{Samsonovich97,Battaglia98,Tsodyks99,Hopfield10}, the activity can be localized in any of the environments, {\em i.e.} the active neurons have neighboring place fields in one map, provided the load $\alpha=L/N$ and the temperature $T$ are not too large. 

\begin{figure}[t]
\includegraphics[width=\linewidth]{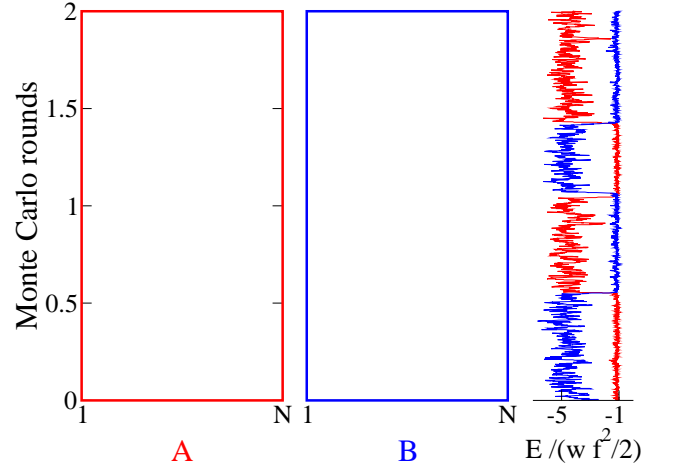}
\caption{(Color online) Time evolution of a network with $N=1000$ neurons, and $L=2$ one-dimensional maps. Each black dot represents an active neuron $\sigma_i$. Panels A and B show the same data, up to a permutation of the neuron indices $i$ to sort place field centers in increasing order in the corresponding map ($x$-axis). Right: contributions to the total energy due to each map, divided by $\frac 12 f^2w$ (absolute value of the PM energy). Parameter values: $T=0.007, f=0.1, w=0.05$. Each MC round includes $10^6$ steps.}
\label{fig:dynamics}
\end{figure}

The purpose of this paper is to study the transitions between different maps, {\em i.e.} how the population activity can abruptly jump from being localized in one map to another one. For definiteness we choose the $N$ place-fields centers to be the nodes $\vec x_i$  of a regular $D$-dimensional cubic lattice. The coupling function in (\ref{LP}) is set to $J(r)= \frac 1N$ if $r<r_c$, and 0 otherwise, where $r$ measures the distance on the grid; the cut-off distance $r_c$ is such that each neuron is connected to its ${w\,N}$ closest neighbors in each map. The fraction of active neurons is fixed to the value $f$. We report in Fig.~\ref{fig:dynamics} the outcome of Monte Carlo simulations of model (\ref{ours}) with $L=2$ maps, referred to as A and B, in dimension $D=1$ (See Supplemental Material at [url] for a simulation in $D=2$ dimensions). The bump of activity diffuses within one map with little deformation, and sporadically jumps from map A to B, and back.  We study below the mechanisms underlying those transitions, remained poorly understood so far.

To capture the typical properties of model (\ref{ours}) we compute its free energy under the constraint that the average activities of neurons whose place fields are centered  in $\vec x$ in map A and in $\vec y$ in map B are equal to, respectively, $\rho^A(\vec x)$ and $\rho^B(\vec y)$. We use the replica method to average the free energy over the random permutations $\pi^\ell$. The outcome, within the replica symmetric hypothesis and for $N\to\infty$, is the free energy per neuron: 
\begin{eqnarray}\label{eq:F}
 \mathcal{F}&=&-\frac12\int\mathrm{d}\vec x\int\mathrm{d}\vec x'\,\rho^A(\vec x) J(|\vec x-\vec x'|) \rho^A(\vec x')  \\ &-&\frac12 \int\mathrm{d}\vec y\int\mathrm{d}\vec y\,'\ \rho^B(\vec y) J(|\vec y-\vec y\,'|) \rho^B(\vec y\,')+\frac{\alpha\beta}2 r \big(f-q\big)\nonumber \\
&+& \int\mathrm{d}\vec x\,\mu^A(\vec x)\rho^A(\vec x)+\int\mathrm{d}\vec y\,\mu^B(\vec y)\rho^B(\vec y)- \eta \, f\nonumber \\
& -&\alpha \sum_{\vec k\ne \vec 0} \bigg[\frac{(q-f^2)\Lambda_{\vec k}}{1-\beta (f-q)\Lambda_{\vec k}}-T\log \big( 1- \beta(f-q)\Lambda_{\vec k}\big) \bigg] \nonumber \\ 
&-&T\int\mathrm{d}\vec x\,\mathrm{d}\vec y\,\mathrm{D}z\log\big(1+ e^{\beta ( \mu^A(\vec x) + \mu^B(\vec y)+ z\sqrt{\alpha\, r} -\eta)}\big)\nonumber \ ,
\end{eqnarray}
where $\beta=1/T$, $\mathrm{D}z\equiv\exp(-z^2/2)/\sqrt{2\pi}$, $\Lambda_{\vec k}\equiv \prod_{\mu=1}^D\sin(k_\mu\,\pi\, w^{1/D})/(\pi k_\mu)$ are the eigenvalues of the $J$ matrix, and the components $k_\mu$ are positive integer numbers. Fields $\mu^A(\vec x)$, $\mu^B(\vec y)$, and parameter $r$ are conjugated to, respectively, the densities $\rho^A(\vec x)$, $\rho^B(\vec y)$, and the Edwards-Anderson overlap $q =\frac 1N \sum _i \overline{\langle \sigma _i\rangle^2}$, where $\langle\cdot\rangle$ denotes the Gibbs average with energy (\ref{ours}) at temperature $T$, and $\overline{(\cdot)}$ is the average over the random permutations. Parameter $\eta$ is chosen to enforce  the fixed-activity constraint: $\int d\vec x \rho^A(\vec x)=\int d\vec y\rho ^B(\vec y)=f$. 

\begin{figure}[b]
\includegraphics[width=\linewidth]{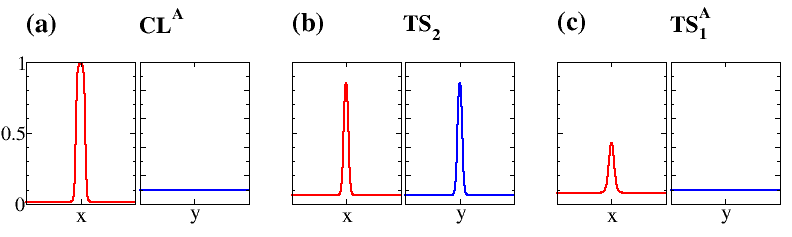}
\caption{(Color online) Activity profiles $\rho^A(x)$ (red, left panels) and $\rho^B(y)$ (blue, right panels) computed from replica theory for {\bf (a)} the phase CL$^A$ ($T=0.006$, {\bf (b)} the transition state TS$_2$ ($T=0.006$), {\bf (c)} the transition state  TS$_1^A$ ($T=0.007$). Profiles attached to CL$^B$ and TS$_1^B$ are obtained from CL$^A$ and TS$_1^A$ by swapping $\rho^A$ and $\rho^B$.  Densities were discretized over $201$ bins. Parameter values: $f=0.1,w=0.05, L=2$ one-dimensional maps.}
\label{fig:shape}
\end{figure}

\begin{figure}[t]
\begin{center}
\includegraphics[width=0.45\linewidth]{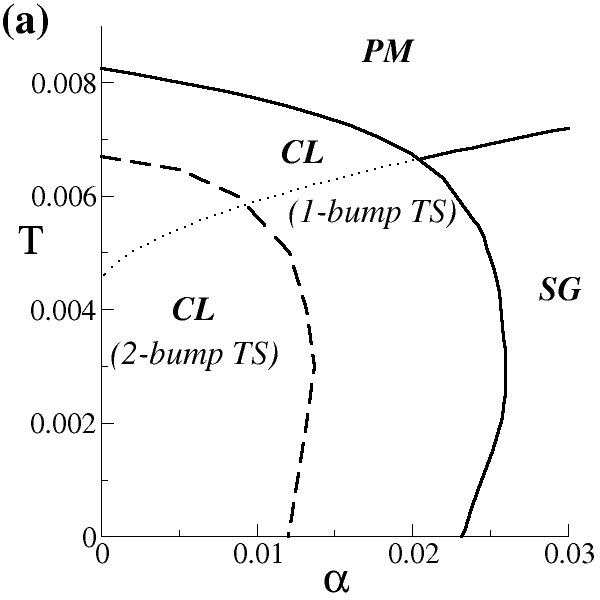}
\hskip .4cm
\includegraphics[width=0.45\linewidth]{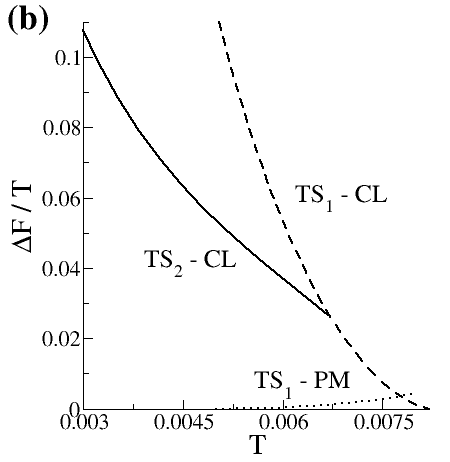}
\end{center}
\caption{{\bf (a)} Region of stability of the CL phase and PM-SG boundary in the $(\alpha,T)$ plane (thick lines); the PM and SG phases coexist with the CL phase, respectively, above and below the dotted line. Dashed line: boundary between 1- and 2-bump transition state (TS) scenarios. {\bf (b)} Theoretical barriers $\Delta {\cal F}/T$ vs. temperature $T$ for $L=2$ environments.  Same parameter values as in Fig.~\ref{fig:shape}.}
\label{fig:phasediag}
\end{figure}

Minimization of ${\cal F}$ allows us to determine the properties of the `clump' phase (CL), in which the activity has a bump-like profile in one map, and is flat ($=f$) in the other one, see Fig.~\ref{fig:shape}(a). Details about the activity profile for a given realization of the maps are shown in Supplemental Material. This CL phase corresponds to the retrieval of one map. The phase diagram in the $\alpha,T$ plane is shown in Fig.~\ref{fig:phasediag}(a), and also includes the paramagnetic (PM) phase at high $T$ and the spin-glass (SG) phase at high $\alpha$, in which no map can be retrieved \cite{Monasson13}. In the PM phase, the neural noise is large enough to wipe out any interaction effect: the average activity $\langle \sigma_i \rangle$ of each neuron coincides with the global average activity, $f$. In the SG phase, activities are non uniform ($\langle \sigma_i\rangle\ne f$) for a given realization of the maps, but reflect the random crosstalk between the maps: they do not code for a well-defined position in any environment. 

\begin{figure}
\includegraphics[width=\linewidth]{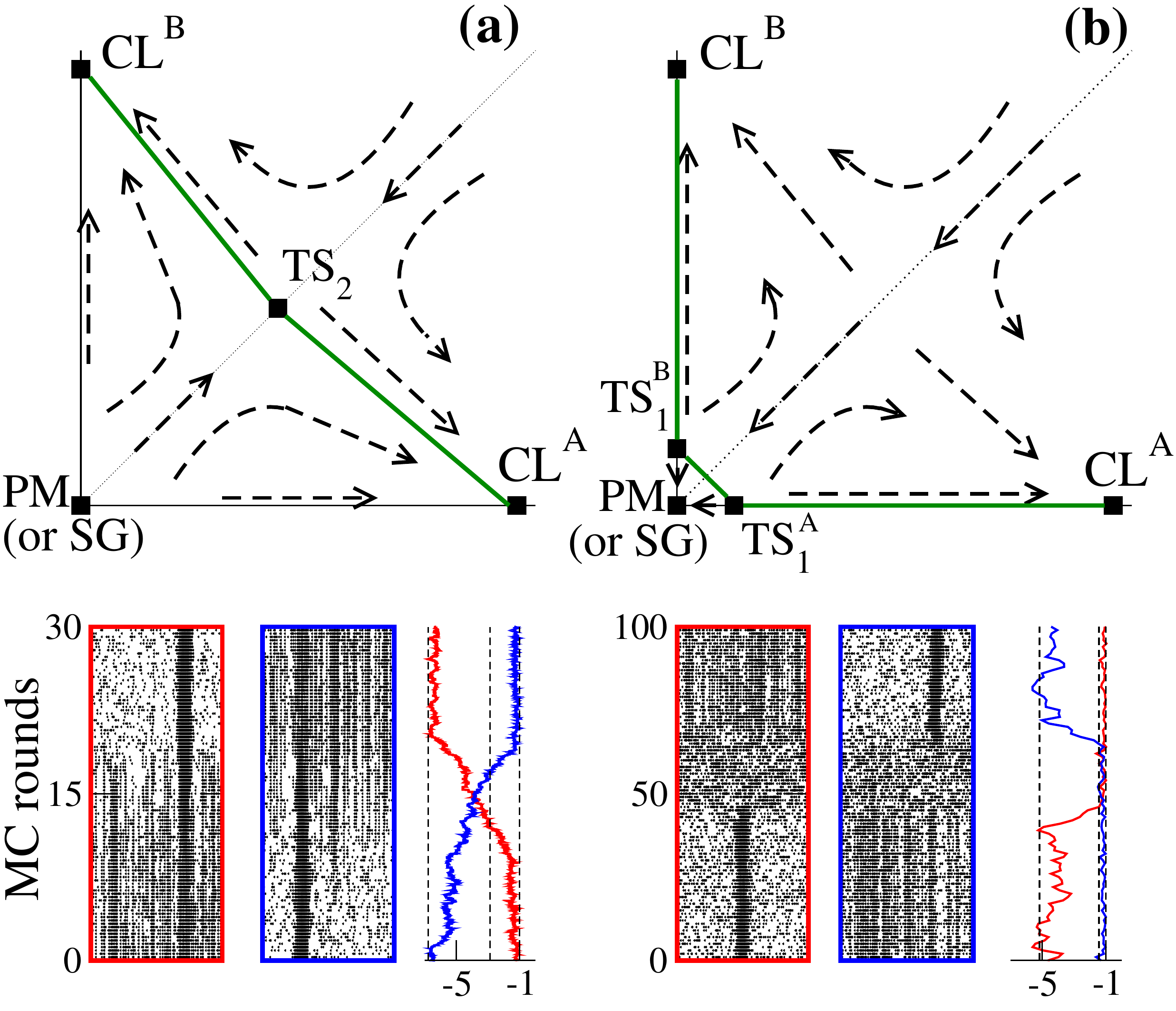}
 \caption{(Color online) Top: representative points of the CL$^A$, CL$^B$, PM or SG phases, and of the saddle-points TS$_2$ (left), TS$_1^A$ and TS$_1^B$ (right). Coordinates correspond to the averaged squared densities in maps A ($x$-axis) and B ($y$-axis), and arrows represent the stability of the phases and saddle-points. The transition pathways are sketched by the green straight lines. Bottom: Monte Carlo simulations illustrating the transition scenarios, and energy contributions due to each map in units of $f^2w/2$ ($x$-axis as in Fig.~\ref{fig:dynamics}); Dotted lines, from left to right: theoretical values of the energies in the CL, TS, PM phases. Parameter values: $N=1000,f=0.1,w=0.05,L=2$, $T=0.006$ (left) and $T=0.007$ (right).}
  \label{fig:scenarios}
 \end{figure}

To understand the transition mechanisms we look for saddle-points of ${\cal F}$, through which the transition pathway connecting CL$^A$ to CL$^B$ will pass with minimal free-energy cost $N\, \Delta {\cal F}$. According to nucleation theory \cite{Langer69} we expect those transition states (TS) to be unstable along the transition pathway, and stable along the other directions. We identify two types of TS, referred to as 1 and 2 respectively, depending on the number of maps in which the activity is localized at the saddle-point (Fig.~\ref{fig:shape}(b)\&(c)). The corresponding transition pathways are schematized in Fig.~\ref{fig:scenarios}. Which type of TS is chosen by the system depends on the values of $\alpha$ and $T$, see Fig.~\ref{fig:phasediag}(a).

\begin{figure}[h!]
\begin{center}
\hskip-.2cm
\includegraphics[width=0.54\linewidth]{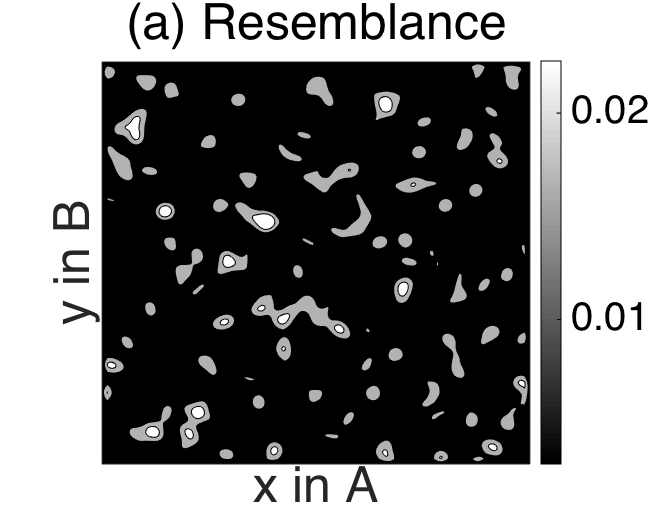}
\hskip -.0cm
\includegraphics[width=.46\linewidth]{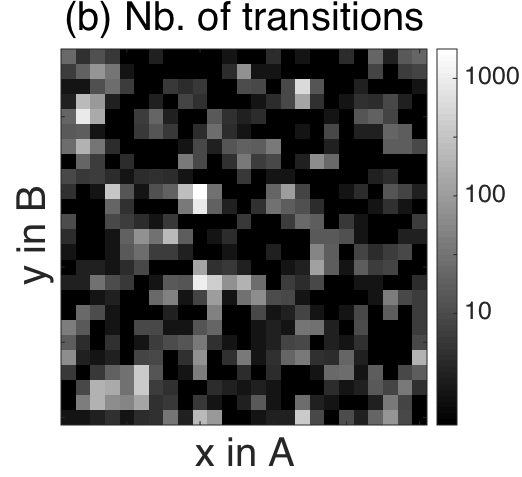}
\end{center}
\vskip -.5cm
\caption{{(\bf a)} Local resemblance of maps (contour lines of Res$_{AB}(x,y)$), and {\bf (b)} number of transitions between positions $x$ in A and $y$ in B observed in Monte Carlo simulation, for two randomly drawn one-dimensional maps A and B. Parameters: $T=0.006,f=0.1,w=0.05,L=2,N=333$. Transitions were counted over $10^6$ rounds of $10N$ MC steps, with starting/ending positions binned in 25 cells in each map.}
\label{fig:positransi}
\end{figure}

At low enough temperature the transition pathway passes through a transition state, TS$_2$, where the activity is equally localized in both maps (Fig.~\ref{fig:shape}(b)). TS$_2$ is a minimum of the free-energy ${\cal F}$ in the symmetric subspace $\rho^A=\rho^B$, and is unstable against one transverse fluctuation mode (Fig.~\ref{fig:scenarios}(a)). The barrier $\Delta {\cal F}$ is the difference between the free-energies of TS$_2$ and CL \cite{Langer69}, and is shown as a function of temperature in the case $L=2$ in Fig.~\ref{fig:phasediag}(b). For a given realization of the maps, the lowest free-energy barrier will be achieved by centering the bumps around positions $\vec x$ in A and $\vec y$ in B, such that the maps are locally similar, {\em i.e.} such that the adjacency matrices of the maps locally coincide. We define the local resemblance between maps A and B at respective positions $\vec x$ and $\vec y$ through
\begin{equation}\label{res}
 \text{Res}_{AB}(\vec x,\vec y) \equiv \frac 1N\sum\limits_{i=1}^N\rho\left(\vec x-\vec x_{\pi^A(i)}\right)\rho\left(\vec y-\vec x _{\pi^B(i)}\right)\ ,
\end{equation}
where $\rho$ denotes the bump profile of TS$_2$ in Fig.~\ref{fig:shape}, common to the two maps. Res$_{AB}(x,y)$ is shown in Fig.~\ref{fig:positransi}(a) for two randomly drawn one-dimensional maps, and compared in Fig.~\ref{fig:positransi}(b) to the number of transitions, starting at position $x$ in A and ending at position $y$ in B (or vice-versa), observed in Monte Carlo simulations. Both quantities are strongly correlated, showing that transitions take place at positions where maps are most similar, as intuited in \cite{McNaughton96}. In addition, those `gateway' positions are energetically favorable, and kinetically trap the activity bump (Fig.~\ref{fig:dynamics}), hence making transitions more likely to happen. 

At high temperatures or loads no saddle-point of ${\cal F}$ localized in two maps can be found. At such temperatures or loads the PM or the SG phase coexists with CL (Fig.~\ref{fig:phasediag}(a)). The PM/SG phase is separated from CL$^A$ by a transition state TS$_1^A$ (Figs.~\ref{fig:shape}(c)~\&~\ref{fig:scenarios}(b)), and from  CL$^B$ by a transition state TS$_1^B$. The transition pathway goes from CL$^A$ to  TS$_1^A$, then to TS$_1^B$ at constant ${\cal F}$ level, before reaching CL$^B$ (Fig.~\ref{fig:scenarios}(b)). The barrier $\Delta {\cal F}$ is given by the difference between the free energies of CL and TS$_1$, and is shown in Fig.~\ref{fig:phasediag}(b). Note that at intermediate $T$ or $\alpha$ values PM or SG are very shallow local minima of ${\cal F}$ (Fig.~\ref{fig:phasediag}(b), dotted line). The system is likely to transiently visit PM or SG from TS$_1^A$ or TS$_1^B$. The activity configurations are then delocalized in both maps, before eventually condensating into the CL phase. Model~(\ref{ours}) defines a dilute ferromagnet with inhomogeneities in the interaction network, see (\ref{jij}). Bumps of neural activity are likely to melt, and TS$_1$-based transitions to take place, where the network is less dense  \cite{Griffiths69}. 

Two-clump (TS$_2$) and one-clump/delocalized (TS$_1$) transition scenarios are observed in simulations as reported in Fig.~\ref{fig:scenarios} and Supplemental Material, Figs.~4, 5\&6, whether the CL phase coexist with the PM (small loads $\alpha$) or the SG (moderate loads) phases, see Fig.~\ref{fig:phasediag}(a). The two scenarios also coexist over a range of temperatures (Fig.~\ref{fig:phasediag}(b)), and may be alternatively observed in finite-size simulations. As an illustration the second transition in Fig.~\ref{fig:dynamics} is of type 1 (at $\simeq1.05\;10^6$ MC steps), while all three other transitions are of type 2. The boundary line along which both scenarios have equal free-energy barriers is shown in Fig.~\ref{fig:phasediag}(a). Simulations confirm that the transition rate increases with $\alpha$ and $T$, and decreases exponentially with $N$ (Supplemental Material, Figs.~1\&2). 

It is interesting to compare the scenarios above to the experiment by K. Jezek \emph{et al.}~\cite{Jezek11}, in which a rat was trained to learn two environments, A and B, differing by their light conditions. The activity of $\simeq 30$ recorded place cells was observed to rapidly change from being typical of environment A to being typical of environment B, or vice-versa, either spontaneously, or as a result of a light switch. During light cue-induced transitions mixed states, correlated with the representative activities of both maps were observed for a few seconds (Figs.~3a\&b, and Supplementary Figs.~6\&7b in \cite{Jezek11}). Spontaneous transitions were also found to take place, in correspondence to mixed states, or to neural configurations seemingly uncorrelated with A and B, see Fig.~3c in \cite{Jezek11}. Those findings are qualitatively compatible with our two transition scenarios. A quantitative comparison of our model with the neural activity in the CA3 and CA1 areas recorded in \cite{Jezek11} will be reported in a forthcoming publication.

Our work could be extended along several lines, e.g. to study the consequences of rhythms, such as the $\simeq 8$~Hz Theta oscillations, believed to be very important for the exploration of the space of neural configurations \cite{Buzsaki11}, and, hence, for transitions \cite{Jezek11,Stella11}. In addition, we have assumed here, for the sake of mathematical tractability, that the coupling matrix in each map was homogeneous (the number of neighbors of each neuron is uniform across the population), a result of perfect exploration and learning of the environment. In reality, imperfect learning, irregularities in the positions and shapes of place fields, and the sparse activity of place cells in CA1 \cite{Thompson89}, and even more so in CA3 will concur to produce heterogeneities in the coupling matrix. Numerical simulations suggest, however, that the mechanisms of transitions we have analytically unveiled in the homogeneous case are unaltered in the presence of heterogeneities (Supplemental Material). Last of all, the notion of space itself could be revisited. The `overdispersion' of the activity of place cells \cite{Fenton98}, its dependence on task and context \cite{Smith06}, ... suggest that place cells code for `positions' in a very high-dimensional space, whose projections onto the physical space are the commonly defined place fields. Extending our study to the case of generic metric spaces could be very interesting, and shed light on the existence of fast transitions between task-related maps \cite{Jackson07} and, more generally, on the attractor hypothesis as a principle governing the activity of the brain. 

\vskip .0cm\noindent
{\bf Acknowledgements.} We are grateful to J.J. Hopfield and K. Jezek for very useful discussions. R.M. acknowledges financial support from the [EU-]FP7 FET OPEN
project Enlightenment 284801 and the CNRS-InphyNiTi project INFERNEUR.

\bibliographystyle{apsrev4-1}

\end{document}